\documentclass[10pt,a4paper]{article}
\usepackage{amsmath}
\usepackage{amssymb}
\usepackage{array}
\usepackage{calc}
\usepackage{longtable}
\usepackage{multirow,booktabs}
\usepackage{cite,mcite}
\usepackage{relsize}
\usepackage{pstricks}
\usepackage{graphicx}
\usepackage{xspace}
\usepackage{units}
\usepackage{afterpage}
\usepackage{placeins}
\usepackage[compat=1.1.0]{tikz-feynman}

\numberwithin{equation}{section}
\usepackage{mciteplus}
\usepackage[pdfborder={0 0 0}]{hyperref}
\usepackage[format=hang,labelfont=bf,hypcap=true]{caption}
\usepackage{subcaption}
\usepackage{sectsty}
\allsectionsfont{\sffamily}
\subsubsectionfont{\mdseries\itshape\large}
\makeatletter
\DeclareRobustCommand*{\bfseries}{%
  \not@math@alphabet\bfseries\mathbf
  \fontseries\bfdefault\selectfont
  \boldmath
}
\makeatother
\let\spreprint\empty
\newcommand{\preprint}[1]{\def\spreprint{\protect#1}}
\let\sinstitute\empty

\makeatletter
\renewcommand{\maketitle}{\begingroup
  \null\thispagestyle{empty}%
    \ifx\spreprint\empty
      \vskip 5ex
    \else
      \flushright\large\spreprint\vskip 2ex
    \fi
    \vskip 5ex
    \flushleft
      {\sffamily\bfseries\huge\@title}\vskip 6ex
      \@author\vskip 2ex
      \ifx\sinstitute\empty
      \else
        {\small\sinstitute}
      \fi
    \vskip 5ex
  \endgroup
}
\makeatother
\renewenvironment{abstract}{\begin{center}
  {\large\sffamily\bfseries Abstract: }
  \begin{minipage}[t]{0.75\textwidth}
}{\end{minipage}\end{center}\vskip 10ex}

\numberwithin{equation}{section}

\newcommand{\bea}{\begin{eqnarray}}
\newcommand{\eea}{\end{eqnarray}}
\newcommand{\beq}{\begin{equation}}
\newcommand{\eeq}{\end{equation}}

\newcommand{\bs}{\begin{split}}
\newcommand{\es}{\end{split}}

\newcommand{\bi}{\begin{itemize}}
\newcommand{\ei}{\end{itemize}}
\newcommand{\bc}{\begin{center}}
\newcommand{\ec}{\end{center}}
\newcommand{\bac}{\begin{array}{c}}
\newcommand{\bacc}{\begin{array}{cc}}
\newcommand{\ea}{\end{array}}

\def\spa#1.#2{\langle#1\,#2\rangle}
\def\spb#1.#2{[#1\,#2]}

\newcommand{\MSbar}{\ensuremath{\overline{\mathrm{MS}}}}

\newcommand{\NLO}{\ensuremath{\mathrm{NLO}}\xspace}

\newcommand{\sla}[1]{\ensuremath{{#1\kern-0.45em/}}}

\newcommand{\eg}{{\itshape e.g.}\xspace}
\newcommand{\ie}{{\itshape i.e.}\xspace}
\newcommand{\cf}{{\itshape cf.}\xspace}
\newcommand{\etc}{{\itshape etc.}\xspace}

\newcommand\lep{L\scalebox{0.8}{EP}\xspace}

\newcommand\LEP{\lep}

\newcommand\FCCee{F\scalebox{0.8}{CC-ee}\xspace}
\newcommand\HERA{H\scalebox{0.8}{ERA}\xspace}
\newcommand\EIC{E\scalebox{0.8}{IC}\xspace}
\newcommand\lepii{L\scalebox{0.8}{EP}~2\xspace}

\newcommand\OPAL{\opal}
\newcommand\opal{O\scalebox{0.8}{PAL}\xspace}

\newcommand\hera{H\scalebox{0.8}{ERA}\xspace}

\newcommand\zeus{Z\scalebox{0.8}{EUS}\xspace}
\newcommand\ZEUS{\zeus}

\newcommand{\MCatNLO}{M\protect\scalebox{0.8}{C}@N\protect\scalebox{0.8}{LO}\xspace}

\newcommand{\Pythia}{P\protect\scalebox{0.8}{YTHIA}\xspace}

\newcommand{\OpenLoops}{O\protect\scalebox{0.8}{PEN}L\protect\scalebox{0.8}{OOPS}\xspace}

\newcommand{\Sherpa}{S\protect\scalebox{0.8}{HERPA}\xspace}
\newcommand{\Comix}{C\protect\scalebox{0.8}{OMIX}\xspace}

\newcommand{\Amegic}{A\protect\scalebox{0.8}{MEGIC}\xspace}
\newcommand{\CSS}{CSS\protect\scalebox{0.8}{HOWER}\xspace}

\newcommand{\Ahadic}{A\protect\scalebox{0.8}{HADIC}\xspace}

\newcommand{\Rivet}{R\protect\scalebox{0.8}{IVET}\xspace}

\usepackage{authblk}
\hypersetup{
  pdfauthor={Stefan Hoeche, Frank Krauss, Peter Meinzinger}
  pdftitle={Resolved Photons in Sherpa},
  pdfkeywords={QCD, Event Generators, Photon PDF, HERA, LEP}
}
\begin{document}
\preprint{FERMILAB-PUB-23-623-T, IPPP/23/59}
\title{Resolved Photons in \protect\Sherpa}
\author[1]{Stefan~H{\"o}che}
\author[2]{Frank~Krauss}
\author[2]{Peter~Meinzinger}
\affil[1]{Fermi National Accelerator Laboratory, Batavia, IL 60510, USA}
\affil[2]{Institute for Particle Physics Phenomenology, Durham University, Durham DH1 3LE, UK}

\maketitle
\begin{abstract}
We present the first complete simulation framework, in the \Sherpa event generator, for resolved photon interactions at next-to leading order accuracy.
It includes photon spectra obtained through the equivalent-photon approximation, parton distribution functions to parametrize the hadronic structure of quasi-real photons, the matching of the parton shower to next-to leading order QCD calculations for resolved photon cross sections, and the modelling of multiple-parton interactions.
We validate our framework against a wide range of photo-production data from \lep and \hera experiments, observing good overall agreement. We identify important future steps relevant for high-quality simulations at the planned Electron-Ion Collider. 
\end{abstract}

\section{Introduction}\label{Sec:Intro}
Photon--\-induced processes provide a rich testing ground for a wide range of physics effects.
This is, on one hand, because photons will couple to any electromagnetically charged particle, resulting in a wide spectrum of accessible final states.
On the other hand, photons have the quantum numbers of (neutral vector) mesons and their wave functions therefore have a sizeable hadronic component which lets them interact strongly, with correspondingly large cross sections.
The production of low--\-multiplicity final states in $\gamma\gamma$ collisions has been observed in many experiments~\cite{BaBar:2018zpn,Ragoni:2021hmr,Klein:2021syg,Tu:2020mvm}, and yields interesting insights into the physics of hadrons and hadron resonances described by effective theories of the strong interactions.  
At increasing centre-\-of-\-mass energies of the colliding photons, new channels open up, and the production of jets has been studied, for example, at \LEP~\cite{OPAL:2003hoh,OPAL:2007jeb,L3:2004ehh,OPAL:1999pnw,OPAL:1998ggd,OPAL:1999peo}.
Similarly, the photo--\-production of various final states, including jets, has been analysed by the \HERA experiments~\cite{ZEUS:2001zoq,ZEUS:2012pcn,ZEUS:1996uid,ZEUS:1998agx,ZEUS:2002wfj,H1:2003xoe,H1:2000kis}.

The results obtained in these experiments have allowed to parameterise the parton content of the hadronic component of quasi-real photons in the form of parton distribution functions (PDFs), for example in the 
GRV~\cite{Glueck1992}, CJK~\cite{Cornet:2002iy,Cornet:2004nb}, SAL~\cite{Slominski2005}, and SaS~\cite{Schuler1995,Schuler1996a} sets\footnote{
    To fix the (non--\-perturbative) inputs for the PDFs at some infrared scale 
    $Q_0^2\approx\mathcal{O}(1\,{\rm GeV}^2)$ different ans\"atze have been chosen. 
    Most of them rely on the vector meson dominance model which assumes that at these scales the hadronic component of the photon wave function behaves like a vector meson, and use the pion PDF with modifications inspired by the quark model.
    The more involved SaS model for the photon structure function is based on the decomposition of the photon wave function $|\gamma\rangle$ into three components: a "bare" photon component $|\gamma_{\rm bare}\rangle$ where the photon interacts indeed as a point--\-particle, a non--\-perturbative component $|\gamma_V\rangle$ where the photon fluctuates into various neutral vector mesons $V$ such as $\rho$, $\omega$, $\phi$, $J/\psi$ \etc, and a perturbative "anomalous component" $|\gamma_{q\bar q}\rangle$ in which the photon fluctuates into a $q\bar q$ pair. 
    In all cases, the PDF at the low scale $Q_0^2$ is translated to PDFs at larger scales through standard QCD evolution, by emitting secondary virtual partons.}
.
In fact, based on these PDFs, satisfying agreement between data and calculations has been achieved, and a complete model for photon structure functions and high--\-energy photon interactions~\cite{Schuler:1996en} has been encoded in \Pythia~\cite{Sjostrand:2006za}.
There, the hard production of QCD final states at large scales, \ie jets, is simulated in the usual way by dressing the hard parton--\-level matrix element with subsequent parton showers, the fragmentation of the resulting partons into hadrons during hadronization, possibly including  an underlying event.  
Recently, the model was extended to also include the perpendicular component of the photon momentum~\cite{Bierlich:2022pfr}.

Pushing for higher accuracy, there have been a few predictions for inclusive jet-production at fixed-order at \HERA~\cite{Klasen:1996it,Frixione:1997ks,Kramer:1998bc,Klasen:1997br,Harris:1997hz} and the \EIC \cite{Guzey:2023syh}, while more attention has recently been paid to exclusive meson production processes and photo-production at heavy-ion collisions~\cite{Chang:2009uj,Eskola:2022vpi,Eskola:2022vaf,Eskola:2023oos,Jones:2015nna,Jones:2013pga}.

Anticipating the increased precision requirements for successfully operating a possible future lepton collider such as \FCCee or the planned electron--\-ion collider, \EIC, motivates to revisit the physics of photon--\-induced processes and to arrive at fully-differential predictions at Next-To-Leading Order (NLO) in QCD perturbation theory. 
We report here on the systematic inclusion of PDFs for quasi--\-real photons into the \Sherpa event generation framework~\cite{Gleisberg:2008ta,Bothmann:2019yzt}, the modelling of multiple-parton interactions, and the extenstion of the calculation to Next-to-Leading Order matched to the parton shower.
The paper is organised as follows.  In Sec.~\ref{Sec:EPA_PDF} we briefly discuss how \Sherpa combines the equivalent photon approximation (EPA) and photon PDFs, with some emphasis on the efficient integration over the resulting phase space, the matching to the parton shower at NLO and the multiple-parton interactions (MPI) model.
In Sec.~\ref{Sec:MC} we will show some first results of full Monte Carlo simulations, comparing results obtained with \Sherpa to data from both the \LEP and \HERA experiments at \MCatNLO accuracy. 
We present comparisons to Leading Order and study the effect of the MPI. 
We summarise our findings in Sec.~\ref{Sec:summary}.

\section{Equivalent Photons and their PDFs}\label{Sec:EPA_PDF}

For the simulation of photo-production events in \Sherpa, we use its existing EPA interface, improved the phase space handling for the initial states for a more efficient integration, and added relevant photon PDFs to \Sherpa's internal PDF interface. 
The resulting code will be publically available as part of the upcoming release of \Sherpa 3.0; in the meantime it can be obtained from the authors upon request.

\subsection{Phase Space Handling}\label{Sec:PS}

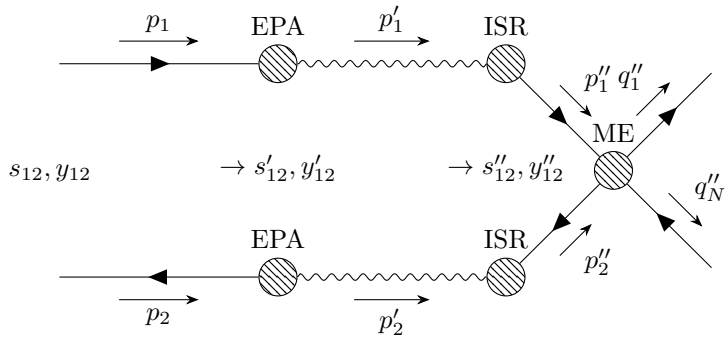
\begin{figure}[h!]
    \centering
    \begin{tikzpicture}[align=center,node distance=4cm]
    \centering
    \begin{feynman}
    \vertex (a) {};
    \vertex[small,blob,label={EPA},right=3cm of a] (b) {};
    \vertex[small,blob,label={ISR},right=3cm of b] (c) {};
    \vertex[small,blob,label={ME},below right=2cm of c] (d) {};
    \vertex[small,blob,label={ISR},below left=2cm of d] (e) {};
    \vertex[small,blob,label={EPA},left=3cm of e] (f) {};
    \vertex[left=3cm of f] (g) {};
    \vertex[above right=2cm of d] (h) {};
    \vertex[below right=2cm of d] (i) {};
    \vertex[label={$s_{12}, y_{12}$},below=1.8cm of a] (j) {};
    \vertex[label={$\rightarrow s'_{12}, y'_{12}$},below=1.8cm of b] (k) {};
    \vertex[label={$\rightarrow s''_{12}, y''_{12}$},below=1.8cm of c] (l) {};

    \diagram*{
        (a) --[large,fermion,momentum={[arrow shorten=0.3]$p_1$}] (b),
        (b) --[large,boson,momentum={[arrow shorten=0.3]$p'_1$}] (c),
        (c) --[large,fermion,momentum={[arrow shorten=0.3]$p''_1$}] (d),
        (d) --[large,fermion,reversed momentum={[arrow shorten=0.3]$p''_2$}] (e),
        (e) --[large,boson,reversed momentum={[arrow shorten=0.3]$p'_2$}] (f),
        (f) --[large,fermion,reversed momentum={[arrow shorten=0.3]$p_2$}] (g), 
        (d) --[large,fermion,momentum={[arrow shorten=0.3]$q''_1$}] (h),
        (d) --[large,anti fermion,momentum={[arrow shorten=0.3]$q''_N$}] (i)
    };
    \end{feynman}
    \end{tikzpicture}
    \caption{Schematic sketch of the phase space mappings between the different steps in the initial states, \ie the Equivalent Photon Approximation (EPA) and the Initial State Radiation (ISR), and the Matrix Element (ME). Each coordinates pair of Mandelstam-$s^{\prime(\prime)}$ and rapidity $y^{\prime(\prime)}$ is randomly sampled and the momenta are calculated as functions of these.}
    \label{fig:ps-sketch}
\end{figure}

In the following we detail the structures for efficient phase space sampling, using the most involved example of doubly resolved photon--photon collisions at lepton colliders, schematically depicted in Fig. \ref{fig:ps-sketch}.
The two incoming leptons have momenta $p_1$ and $p_2$, and a (beam) c.m.-system characterised by the c.m.-energy squared $s_{12}$ and its rapidity $y_{12}$ in the lab system.
The momenta of the photons emitted by the leptons, $p'_1$ and $p'_2$, create a (photon) c.m.-system characterised by its c.m.-energy squared $s'_{12}$ and rapidity $y'_{12}$ with respect to the beam system.  
The partonic structure of the photons, as described by the PDFs, results in two partons with momenta $p''_1$ and $p''_2$ to finally enter the hard process which will result in final state particles with momenta $q''_i$. The hard scattering is characterised by a c.m.-energy of $s''_{12}$ and a rapidity $y''_{12}$ w.r.t.\ the photon system.
This structure requires two nested integrations for the two successive "initial states" (photons and partons): 
first an integration over $s'_{12}$ and $y'_{12}$, with factors given by the EPA spectra, and then an integration over $s''_{12}$ and $y''_{12}$, with factors given by the PDFs, before adding the integration over the final state phase space over the outgoing momenta $q''_i$.
Efficient integration over this complex phase space in \Sherpa is facilitated through the multi-channel method~\cite{Kleiss:1994qy} with automatically generated integration channels that map out intrinsic structures such as $s$-channel resonances \etc

After the successful generation of a phase space point, the corresponding weight is calculated, given by the factors stemming from the EPA photon spectra and the PDF weights. 

\subsection{Equivalent Photon Approximation}\label{Sec:EPA}

The equivalent photon approximation encoded in the Weizsäcker-Williams formula~\cite{vonWeizsacker:1934sx,Williams:1934ad,Budnev:1974de} is based on the observation that quasi-virtual photons can be approximated through real photons for small virtualities $Q^2 < \Lambda_\text{cut}^2$. 
As photo-production events are dominated by the interaction of low-virtuality photons, the differential cross-section can be substituted by $\text{d}\sigma_{e X} = \sigma_{\gamma X}(Q^2=0) \text{d}n$. 
\Sherpa uses an improved version of the spectrum, following~\cite{Frixione:1993yw}, which introduces the term proportional to $m_e^2$ to the spectrum, and the photons are assumed to be collinear to the electron beam. 
The dependence of the photon spectrum on the photon virtuality is integrated out. 
This results in the following spectrum for electrons: 
\begin{equation}
    \mathrm{d}n = \frac{\alpha_\mathrm{em}}{2 \pi} \frac{\mathrm{d}x}{x}
                \left[ \left( 1 + (1 - x)^2 \right) \log \left( \frac{Q^2_\mathrm{max}}{Q^2_\mathrm{min}} \right) -
                2 m_e^2 x^2 \left( \frac{1}{Q^2_\mathrm{min}} - \frac{1}{Q^2_\mathrm{max}} \right) \right]
\end{equation}
Here, $x$ denotes the ratio of photon to electron energy, $\frac{E_\gamma}{E_e}$, and $\alpha_\mathrm{em}$ is the electromagnetic coupling constant. 
$Q^2_\mathrm{max}$ and $Q^2_\mathrm{min}$ denote the maximal and minimal photon virtuality, where the latter can be calculated from kinematic restrictions and is given by
\begin{equation}
    Q^2_\mathrm{min} = \frac{m_e^2 x^2}{1 - x} \, .
\end{equation}
The maximal virtuality is given by the experimental setup and the maximal deflection angle of the electron, $\theta_\mathrm{max}$, below which the hard process is still considered to be photon-induced. It is given by
\begin{equation}
    Q^2_\mathrm{max} = \mathrm{min} \left( Q^2_\mathrm{min} + E_e^2 (1 - x) \theta_\mathrm{max}^2, Q^2_\mathrm{max,fixed} \right) \,.
\end{equation}
Defaults choices are $\theta_\mathrm{max} = 0.3$ and the maximum virtuality to $Q^2_\mathrm{max,fixed} = 3\ \mathrm{GeV}^2$, but can be overwritten by the user, cf.\ the \Sherpa manual~\cite{sherpa-manual}.

\subsection{Photon PDFs}\label{Sec:PDFs}
To facilitate a comparison over different parameterisations, four PDF libraries have been included in \Sherpa, see Table~\ref{tab:pdfs} for a summary. 
\begin{table}[!ht]
    \begin{center}
        \begin{tabular}{|l||c|c|c|c|c|c|}
        \hline
        Name & \# sets & Virtual? & NLO? & \# flavours & $x$-range & $\mu_F^2$-range \\
        \hline\hline
        $\vphantom{\frac{|^|}{|_|}}$ GRV~\cite{Glueck1992} & 2 & No & Yes & 5 & $[10^{-5}, 1]$ & $[0.25, 10^6]$ \\
        $\vphantom{\frac{|^|}{|_|}}$ SAL~\cite{Slominski2005} & 1 & No & Yes & 6 & $[10^{-5}, 0.9999]$ & $[2, 8 \cdot 10^4]$ \\
        $\vphantom{\frac{|^|}{|_|}}$ SaS~\cite{Schuler1995,Schuler1996a} & 4 & Yes & Yes & 6 & $[10^{-5}, 1]$ & $[0.25, 10^6]$ \\
        $\vphantom{\frac{|^|}{|_|}}$ CJK~\cite{Cornet:2002iy,Cornet:2004nb} & 4 & No & Yes & 5 & $[10^{-5}, 1]$ & $[0.25, 2 \cdot 10^5]$ \\
        \hline
        \end{tabular}
    \parbox{0.8\textwidth}{\caption{
        Photon PDF libraries included in \Sherpa and their properties. }
        \label{tab:pdfs}
    }
    \end{center}
\end{table}

Currently, all PDFs are evaluated at virtuality $Q^2 = 0$; the extension to virtual photons, taking also into account longitudinal polarisations, will be introduced in a later release. 
The extraction of a parton from the photon is complemented by the corresponding treatment of the remaining partons. 
For this a similar procedure is applied as in the remnant construction for a hadron, however, with a few simplifications. 
As there are no valence quarks present in the photon, the remnant is constructed as the anti-particle of the hard-scattering quark. In the case of the gluons, a quark-antiquark pair is constructed from one of the light quark generations. 
In the following processing, the photon remnants is treated analogously to a hadron remnant, \ie its longitudinal momentum is distributed among the partons according to the kinematics, and their transverse momenta, given by some non-perturbative intrinsic $k_T$ distribution, compensate each other. 

\subsection{NLO and matching to the parton shower}\label{Sec:Matching}

The procedure for a \NLO calculation in photo-production has been presented in \cite{Frixione:1997ma}. 
We will follow the same line of arguments for the matching to the parton shower, summarising the procedure in the following paragraphs. 
In contrast to the more familar case of hadronic PDFs, the evolution in the photon PDFs has an additional "source" term, related to the photon splitting into a quark--\-anti-quark pair, which is proportional to $\alpha_\mathrm{em}$. 
This means that the evolution of the parton distributions $f_i^{(\gamma)}$ for parton $i$ in the photon follow~\cite{Frixione:1997ma}
\begin{equation}
    \frac{\partial f_i^{(\gamma)}}{\partial \mathrm{log} \mu^2} = \frac{\alpha_\mathrm{em}}{2 \pi} P_{i\gamma} + \frac{\alpha_S}{2 \pi} \sum_j P_{ij} \otimes f_j^{(\gamma)}\,.
\end{equation}
The first term on the r.h.s.\ of the equation above gives rise to the "anomalous component" mentioned before.

At Next-to-Leading-Order, there will be additional collinear divergences in the real-corrections, stemming from the photon splitting. 
These divergences appear only in the direct-photon component, but are cancelled against the corresponding term in the photon PDF evolution. 

To ensure an exact cancellation between these terms, the PDF has to use the same factorisation scheme as the subtraction scheme, which in the case of \Sherpa is the \MSbar{} scheme. 
This reduces the number of possible PDFs sets directly available for the NLO calculation in \Sherpa down to SAS1M and SAS2M from the SaS library. 
We note that additional PDFs can be made available easily by adding the respective factorization scheme dependent correction terms~\cite{Catani:1996vz}.
Apart from this subtlety in the choice of PDFs both NLO calculations and full simulations proceed in full analogy to the more familiar case of, {\it e.g.},  proton--\-proton collisions.

\section{Validation of the framework}\label{Sec:MC}

We will now turn to the validation of our implementation, comparing results at \MCatNLO precision with photo-production data from the \OPAL and \zeus experiments. 
For each collider set-up and energy we generated samples of $5 \cdot 10^6$ events at \MCatNLO accuracy.
For the calculation of matrix elements we used \Amegic~\cite{Krauss:2001iv} and \Comix~\cite{Gleisberg:2008fv} for the tree-level matrix elements and subtraction terms~\cite{Gleisberg:2007md} and \OpenLoops~\cite{Buccioni:2019sur} for the one-loop matrix elements.
We added the \CSS parton shower~\cite{Schumann:2007mg} for the jet evolution, matched with the \MCatNLO method~\cite{Frixione:2002ik, Frixione:2003ei} as implemented in \Sherpa~\cite{Hoeche:2011fd}.
Underlying event effects have been included through an implementation of the Sjostrand-van Zijl model~\cite{Sjostrand:1987su,Schuler:1993wr} within \Sherpa~\footnote{For details and a future tune to data we refer the reader to a forthcoming \Sherpa manual.}, and the partons were hadronized with the cluster fragmentation of \Ahadic~\cite{Chahal:2022rid}. 
We consistently used the current default value for $\alpha_S = 0.118$ with three-loop running. 
As described in \ref{Sec:Matching}, the event generation must currently use PDFs based on the \MSbar\ scheme, so the resolved-photon predictions were generated with, and averaged over, the SAS1M and SAS2M PDF sets.
The factorisation scale and the renormalisation scale were both set to $\mu_F = \mu_R = H_T/2$ and we kept the 7-point variation for the scale uncertainty estimate.

We used the \Rivet~\cite{Bierlich:2019rhm} framework with the existing analyses implementing~\cite{OPAL:2003hoh,OPAL:2007jeb,L3:2004ehh,ZEUS:2001zoq,ZEUS:2012pcn}. 
For each experiment, different components of the cross-section have to be summed over, for example,
\begin{equation}
    \sigma_\text{tot} = \sigma_{\gamma \gamma} + \sigma_{j \gamma} + \sigma_{\gamma j} + \sigma_{jj}
\end{equation} 
for \LEP and 
\begin{equation}
\sigma_\text{tot} = \sigma_{\gamma j} + \sigma_{jj}
\end{equation}
for \HERA, where in both cases $j$ denotes a photon or proton resolved through a PDF.

\subsection{Comparison with \LEP data}\label{Sec:LEP}

The \OPAL analysis of photon-\-induced di-jet production at 198 GeV c.m.-energy~\cite{OPAL:2003hoh} offers the most differential observables, and we use it as the primary reference for our validation of photo-production at \lepii. 
To comply with experimental cuts, the $k_T$ algorithm with a jet radius of $R = 1.0$ is used and cuts of $E_T > 4.5\ (2.5)$ GeV for the leading (subleading) jet are imposed.  
\begin{equation}
    \label{Eq:xgamma}
    x_\gamma^\pm = \displaystyle{
    \frac{\sum\limits_{j=1,2} E^{(j)}\pm p_z^{(j)}}{\sum\limits_{i\in{\rm hfs}} E^{(i)}\pm p_z^{(i)}}}\,,
\end{equation}
are used in this analysis to disentangle direct, singly, and doubly resolved production. 
In their definition, the sum in the numerator is over the two jets, and the sum in the denominator is over all hadronic final state particles, thereby distilling the energy-fraction the jets have w.r.t.\ the overall photon energies.  
\begin{figure}[h!]
    \centering
    \begin{tabular}{cc}
        \includegraphics[width=.4\linewidth]{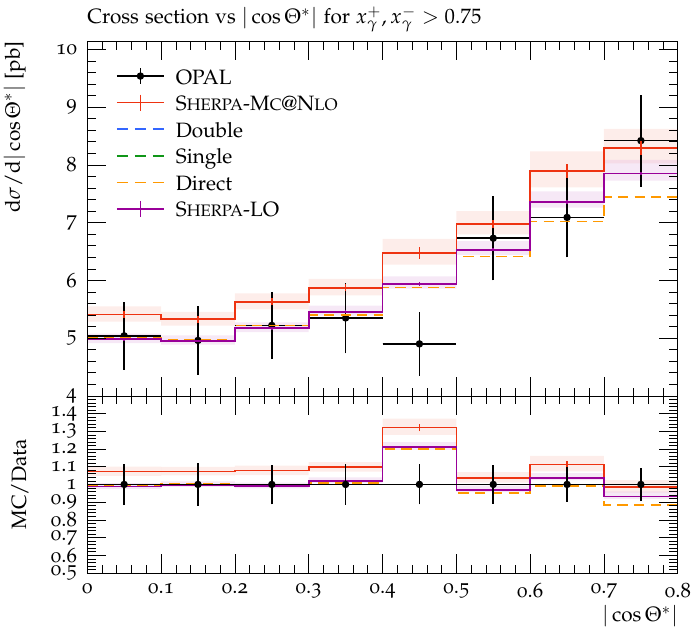} &
        \includegraphics[width=.4\linewidth]{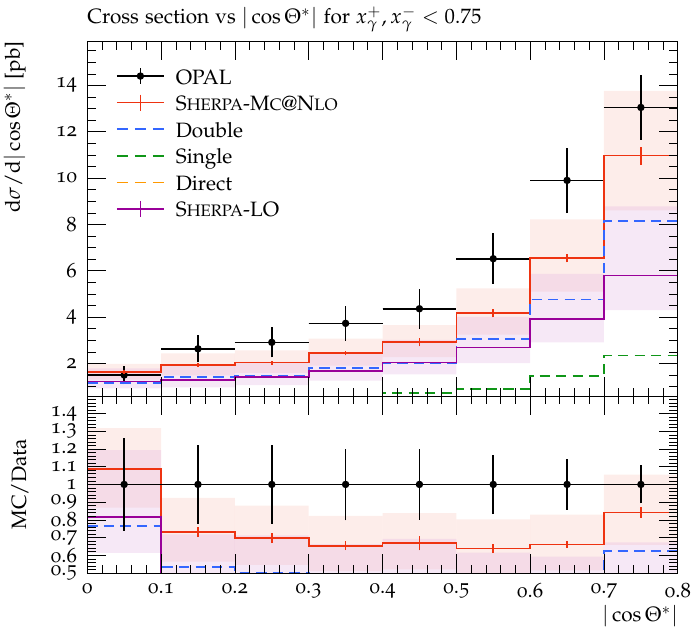} \\
        \includegraphics[width=.4\linewidth]{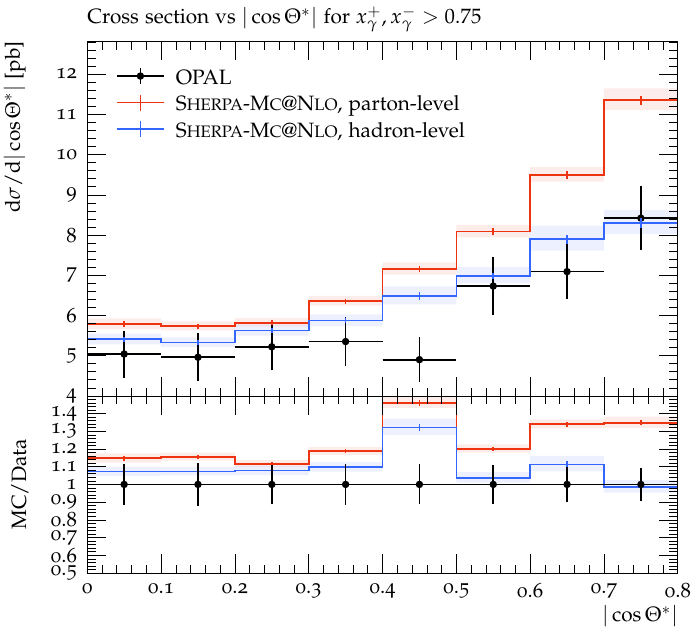} &
        \includegraphics[width=.4\linewidth]{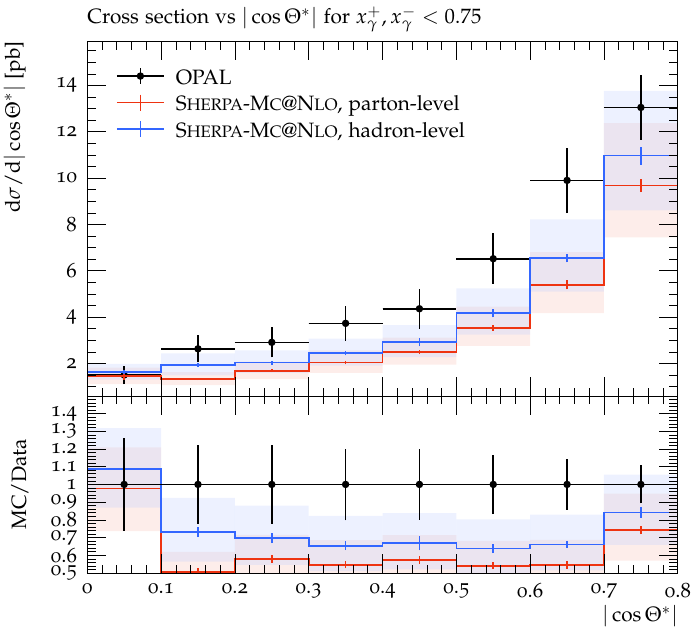}
    \end{tabular}
    \caption{Differential dijet inclusive cross section with respect to $\cos\theta^*$, defined in Eq.~(\ref{Eq:Dijet_kin}), comparing results of our \protect\Sherpa \MCatNLO simulation at hadron-level with the LO simulation (top row) and with simulations at parton-level (bottom row) and with data from \protect\OPAL at an $e^-e^+$ c.m.-energy of 198 GeV~\protect\cite{OPAL:2003hoh}.  
    In the left and right panels the requirement $x^{\rm obs}_\gamma<0.75$ and $x_\gamma^{\rm obs}>0.75$ are applied and enhance resolved and direct contributions, respectively.}
    \label{fig:nlo-lep1-costheta}
\end{figure}
This is exemplified by, \eg, the distribution of events in
\begin{equation}
    \label{Eq:costheta}
    \cos\Theta^* = \tanh\frac{\eta_{1}-\eta_2}{2}\,,
\end{equation}
an approximation of the angle between the two jets, and exhibited in Fig.\ ~\ref{fig:nlo-lep1-costheta}.  
Apart from satisfying agreement with data, a few things are worth noticing here, which we will continue to observe also in the following:
For $x_\gamma^\pm > 0.75$ the direct component dominates by about 1.5 orders of magnitude, with only small scale uncertainties, indicated by the pink band. 
Conversely, for $x_\gamma^\pm < 0.75$ the doubly-resolved component dominates with a significantly larger scale uncertainty, which, in this case, also includes factorization scale uncertainties.  
Intuitively, the singly-resolved component in both case ranges between the two other components.
In addition we observe that hadronization effects reduce the cross section in the unresolved domanin, while the combination of hadronization and multiple parton scattering increases it in the doubly-resolved regime.  
The visible effect in the latter suggests that a careful retuning of the MPIs may further improve agreement with data.

\begin{figure}[!ht]
    \centering
    \begin{tabular}{ccc}
        \includegraphics[width=.3\linewidth]{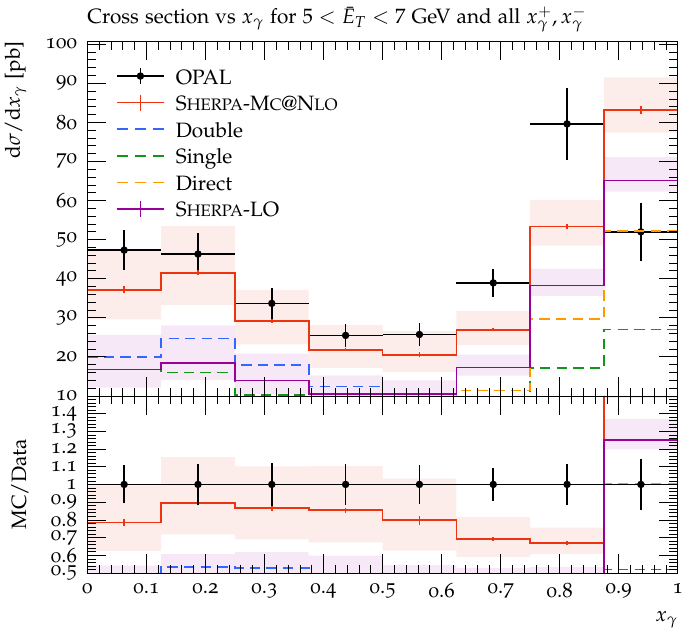} &
        \includegraphics[width=.3\linewidth]{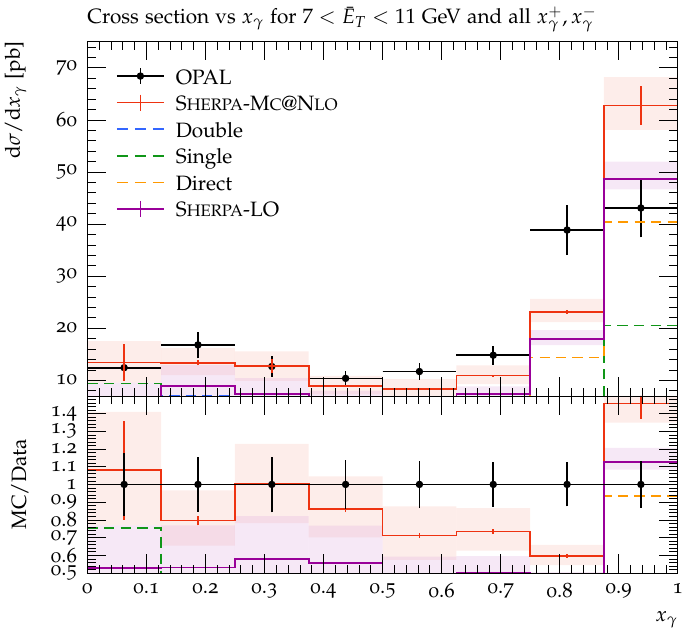} &
        \includegraphics[width=.3\linewidth]{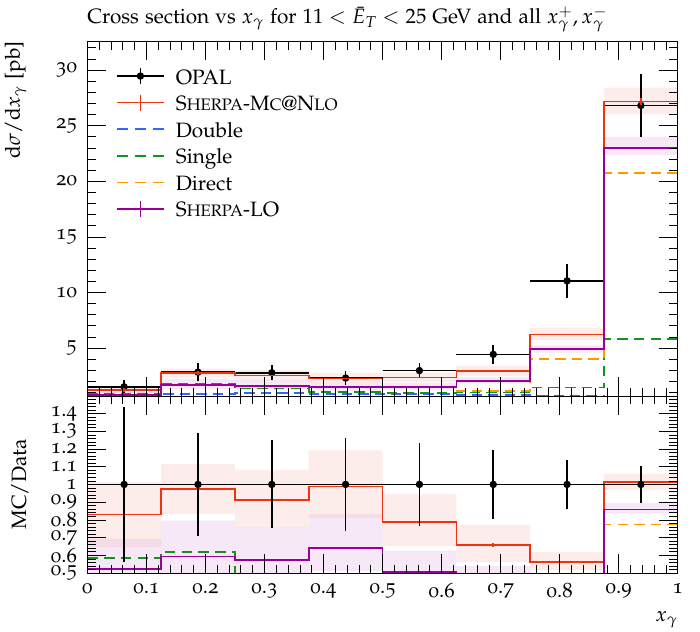}
    \end{tabular}
    \caption{Distributions $x_\gamma^\pm$, collectively denoted as $x_\gamma$ in different bins of average transverse jet energy: $\bar E_T\in [5\,{\rm GeV}, 7\,{\rm GeV}]$ (left), $\bar E_T\in [7\,{\rm GeV}, 11\,{\rm GeV}]$ (middle), $\bar E_T\in [11\,{\rm GeV}, 25\,{\rm GeV}]$ (right).
    Results of the \protect\Sherpa simulation with \MCatNLO accuracy are compared with results at LO and with data from \protect\OPAL at an $e^-e^+$ c.m.-energy of 198 GeV~\protect\cite{OPAL:2003hoh}.}
    \label{fig:nlo-lep1-xgamma}
\end{figure}
We report that distributions in $x_\gamma^{\rm obs}$ for three different $\bar E_T$ experience a significant improvement in shape when going from Leading to Next-to-Leading Order, \cf Fig.\ ~\ref{fig:nlo-lep1-xgamma}. 
However, in the transition region between doubly resolved to unresolved events, we notice a clear difference in shape: 
While for $x_\gamma^{\rm obs}<0.6-0.7$ the prediction is relatively flat below the data, the underprediction at around $x_\gamma^{\rm obs} \approx 0.8$ persists at NLO. 
Apart from possibly insufficient photon PDFs -- a point we will elucidate below -- there are a number of possible explanations:
First of all, as before, a retuning of MPIs may come to the rescue and fill up the gap.  
\begin{figure}[h!]
    \centering
    \begin{tabular}{ccc}
        \includegraphics[width=.3\linewidth]{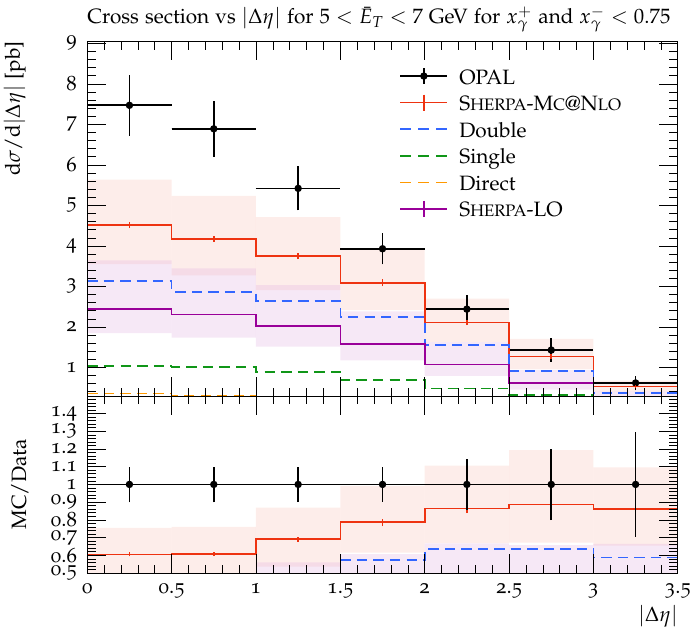} &
        \includegraphics[width=.3\linewidth]{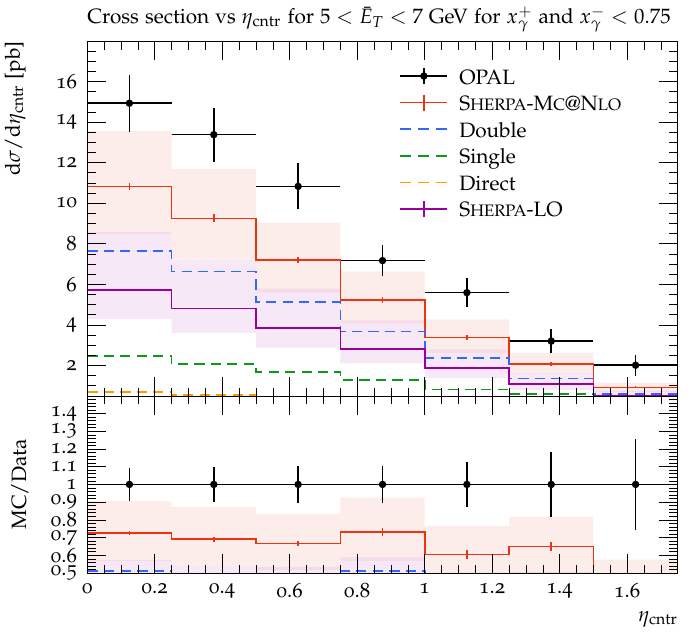} &
        \includegraphics[width=.3\linewidth]{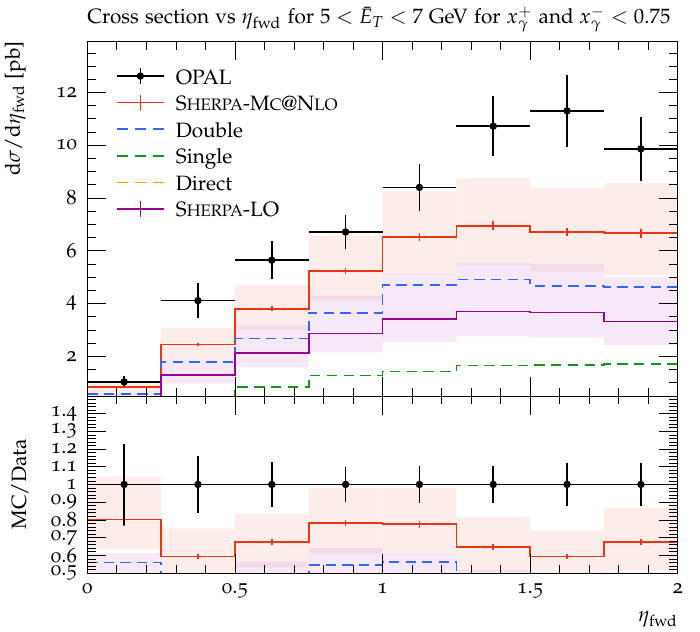} 
    \end{tabular}
    \caption{Distributions of $|\Delta\eta|$ (left), $|\eta_{\rm cntr}|$ (middle), and $|\eta_{\rm fwd}|$ (right), comparing \MCatNLO and LO. 
    Results of the \protect\Sherpa simulation are compared with results from \protect\OPAL at an $e^-e^+$ c.m.-energy of 198 GeV~\protect\cite{OPAL:2003hoh}.}
    \label{fig:nlo-lep1_pl_mpi}
\end{figure}
Secondly, this drop around $x_\gamma\approx 0.7$ could be attributed to the missing QED splitting kernel in the evolution of the parton shower. 
Including this term would impact the backwards evolution of the photonic initial state radiation leading to a photon being reconstructed as the initial state also in the case of a resolved process. 
This again would lead to fewer radiation being generated, therefore shifting the distribution of the resolved process towards larger $x_\gamma$ values. 
The inclusion of this term in the evolution of the initial state showering is left for future work. 
Finally, we should stress that our singly resolved events are described by the $2\to 2$ scattering of on-shell photons with partons from the resolved photons, an approximation which is probably not entirely correct as virtual photons would lead to a DIS-like scattering of the resolved photon, thereby inducing a somewhat different kinematics and scale choices.  

Fig.\ \ref{fig:nlo-lep1_pl_mpi} shows distributions of jet pseudo-rapidities and their differences.  
Again, the overall shape of the prediction is improved and the lowered NLO cross-section is countered by the inclusion of Multiple-Parton Interactions (MPIs).  

\subsection{Comparison with \HERA data}\label{Sec:HERA}
For the further validation of our implementation in electron--\-proton collisions we mainly rely on \ZEUS data~\cite{ZEUS:2012pcn} taken at \HERA Run 2.
The kinematic cuts on the final states in the hard matrix element calculation were chosen to be a minimal transverse momentum of $p_T > 11 (8)$ GeV for the (sub-)leading jets using the $k_T$ clustering algorithm with radius $R = 1.0$ to safely capture the phase space cuts of $p_T^{(1)}>14$ GeV and $p_T^{(2)}>11$ GeV used in the analysis of the \protect\ZEUS Run-1 data~\protect\cite{ZEUS:2001zoq}, even after taking into account MPIs. For the \protect\ZEUS Run-2 data~\protect\cite{ZEUS:2012pcn}, we chose $p_T>13$ GeV to comply with the experimental cut of $E_T>17$ GeV on the leading jet, with otherwise same settings. 
\begin{figure}[hp!]
  \centering
  \begin{tabular}{cc}
      \includegraphics[width=.4\linewidth]{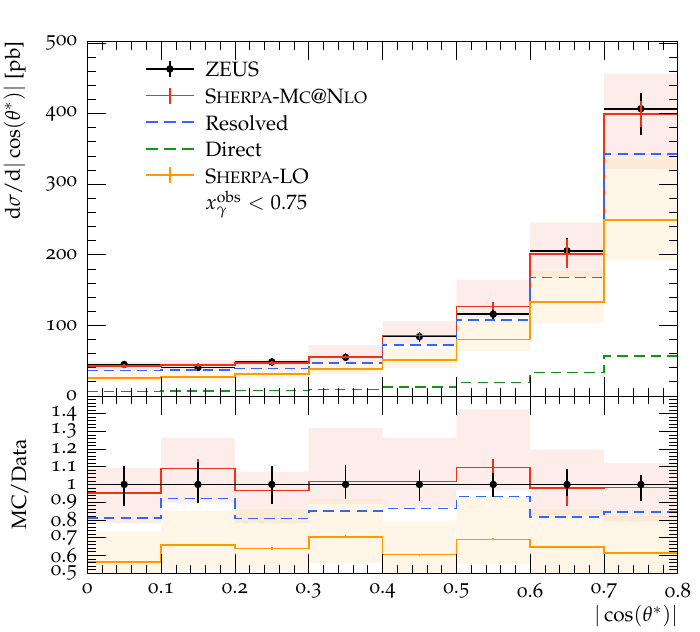} &
      \includegraphics[width=.4\linewidth]{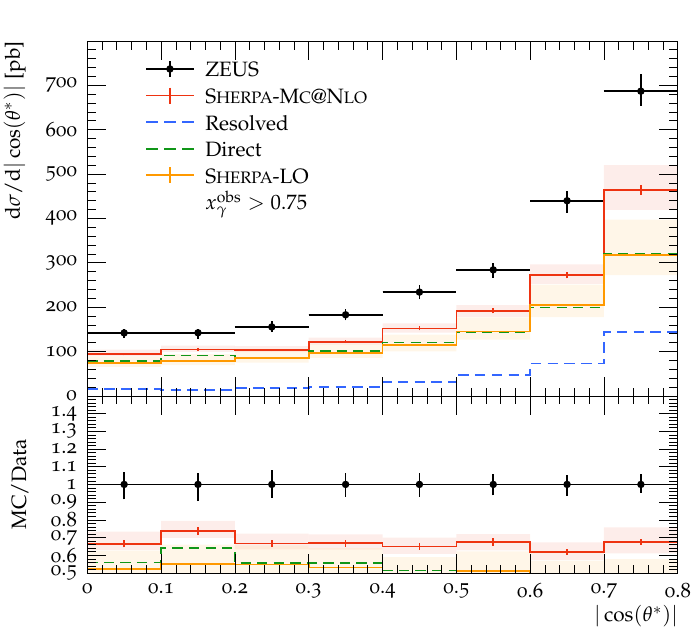} 
    \end{tabular}
    \caption{Differential dijet inclusive cross section with respect to $\cos\theta^*$ for $x^{\rm obs}_\gamma<0.75$ (left) and $x_\gamma^{\rm obs}>0.75$ (right), comparing results of our \protect\Sherpa \MCatNLO simulation with \protect\ZEUS Run-1 data~\protect\cite{ZEUS:2001zoq}.  }
    \label{fig:nlo-zeus-2002a}
\end{figure}
We use the same scale setting algorithm as before, and we also evaluate the theory uncertainties as combination of scale and PDF uncertainties like in the case of electron--\-positron colliders in the section above.
Defining, in analogy to the case at lepton colliders above, Eqs.~(\ref{Eq:xgamma}) and~(\ref{Eq:costheta}) , respectively,
\begin{equation}
\label{Eq:Dijet_kin}
        x_\gamma^{\rm obs} = \frac{E_T^{(1)} e^{-\eta^{(1)}}+ E_T^{(2)} e^{-\eta^{(2)}}}{2yE_e}
        \;\;\;\mbox{\rm and}\;\;\;
        \cos\theta^* = \tanh\frac{\eta^{(1)}-\eta^{(2)}}{2}
        \,.
\end{equation}
Here $^{(1,2)}$ labels the leading and sub-leading jet, $E_e$ is the lepton energy, and $y$ is the energy fraction of the photon w.r.t.\ the lepton.
As before we observe that the $x_\gamma$ is excellently suited to disentangle unresolved and resolved photon interactions, \cf Fig.\ \ref{fig:nlo-zeus-2002a}.
Again we observe satisfactory agreement with data, indicating that the summation over the different components is correct.
Interestingly, as suggested by the right panel of the figure, it appears as if the direct component is not sufficient to fully recover the experimental cross section.
Possible explanations, as before, are related to the missing "anomalous" $\gamma\to q\bar{q}$ splitting in the backwards evolution, or, possibly more relevant here, a failure of the strictly on-shell approximation of the incident photons inherent to the treatment through EPA. 

\begin{figure}[h!]
    \centering
    \begin{tabular}{cc}
        \includegraphics[width=.4\linewidth]{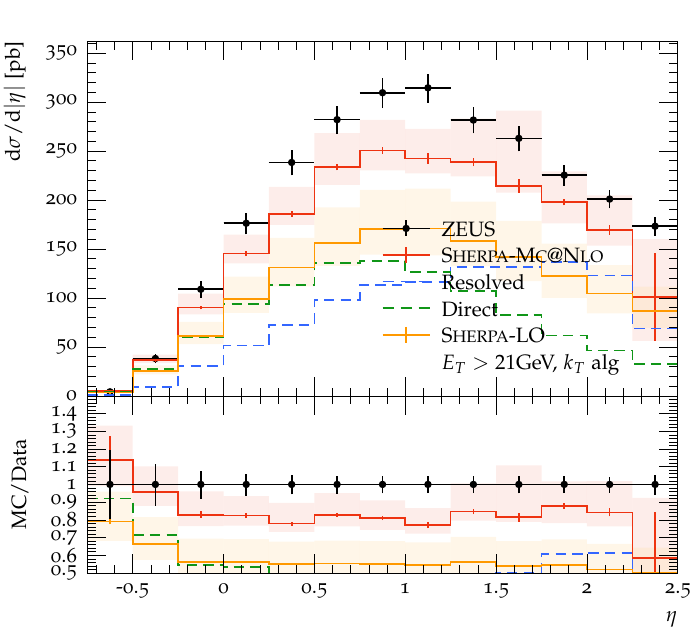} &
        \includegraphics[width=.4\linewidth]{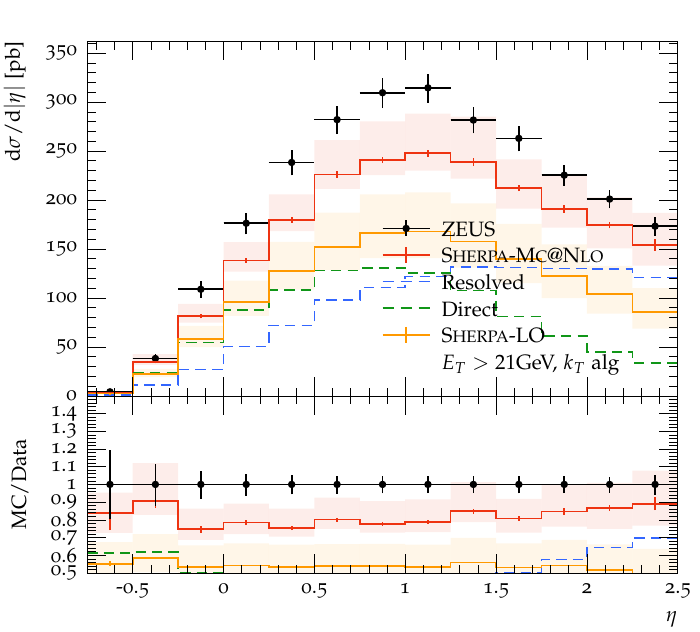}
    \end{tabular}
    \caption{Differential single-jet inclusive photo-production cross section with respect to the pseudo-rapidity of the leading jet, $\frac{\mathrm{d} \sigma}{\mathrm{d} \eta}$, comparing results of our \protect\Sherpa simulation at parton (left) and hadron-level (right) with \protect\ZEUS Run-2 data~\protect\cite{ZEUS:2012pcn}.}
    \label{fig:nlo-zeus-2012-hl-pl}
\end{figure}
\begin{figure}[h!]
    \centering
    \begin{tabular}{cc}
        \includegraphics[width=.4\linewidth]{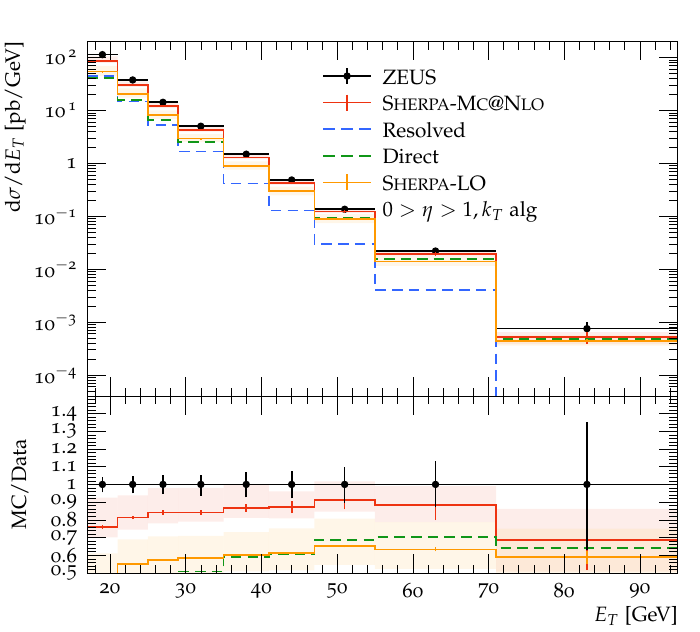} &
        \includegraphics[width=.4\linewidth]{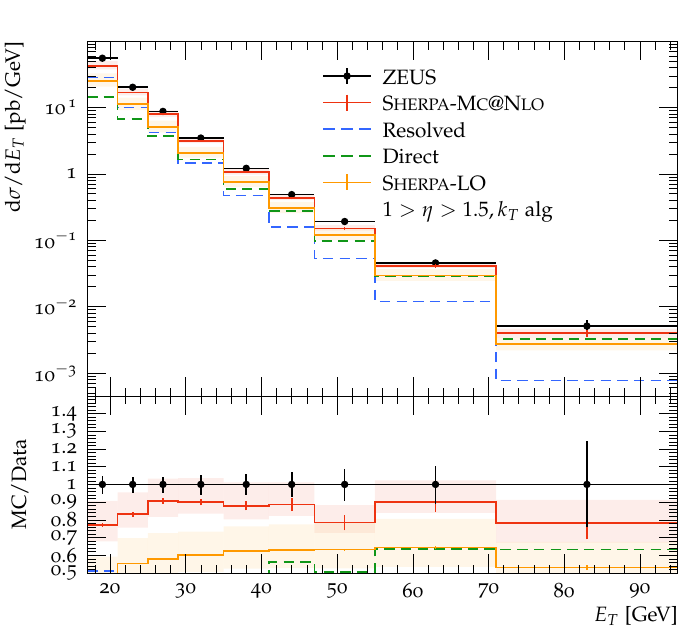} \\
        \includegraphics[width=.4\linewidth]{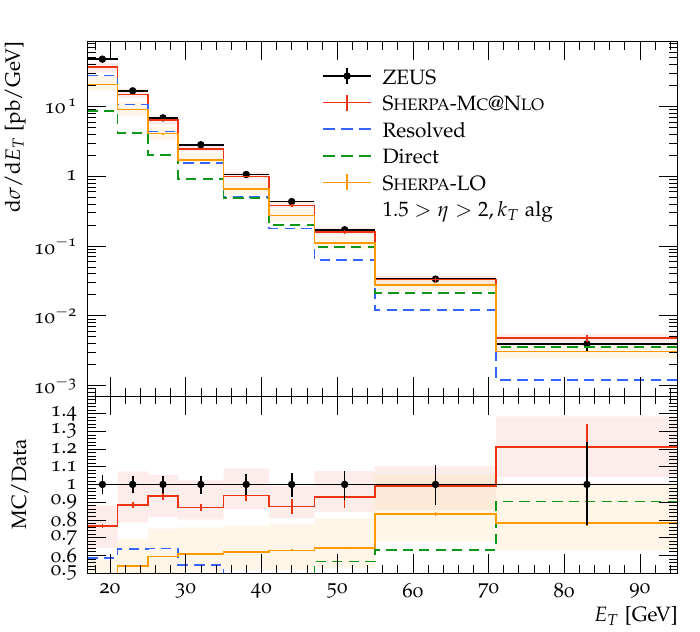} &
        \includegraphics[width=.4\linewidth]{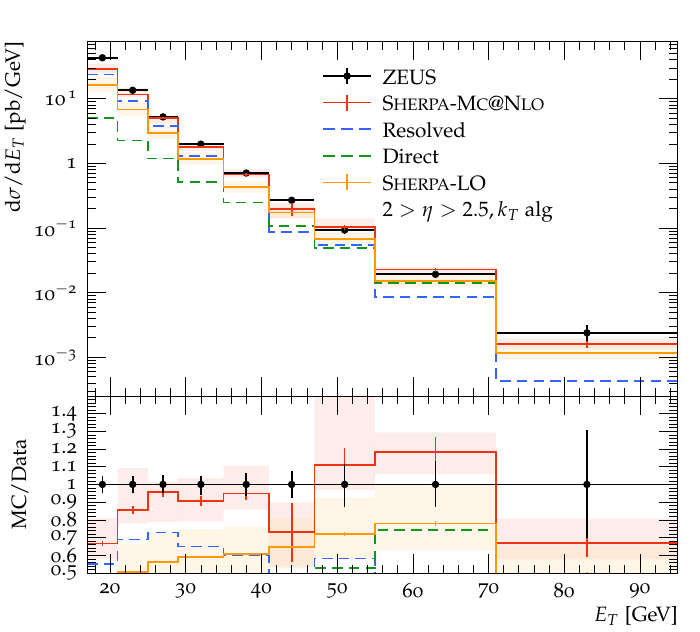}
    \end{tabular}
    \caption{Single-jet inclusive transverse energy spectra for the leading jet in different pseudo-rapidity bins of the $k_T$-jets: $0<\eta<1$ (top left), $1<\eta<1.5$ (top right), $1.5<\eta<2$ (bottom left), and $2<\eta<2.5$ (bottom right), comparing \Sherpa \MCatNLO results with \protect\ZEUS Run 2 data~\protect\cite{ZEUS:2012pcn}.}
    \label{fig:nlo-zeus-2012_2}
\end{figure}
With this caveat in mind we will now turn into a more differential analysis of QCD final states in photo-production at \HERA.
In Fig. \ref{fig:nlo-zeus-2012-hl-pl} we compare the parton- against the hadron-level results and as previously observed, the shape improves significantly through the combined effect of hadronization and MPIs.  
This is most visible in the phase space of $\eta < 0$, rendering the distribution of the simulation data flat compared to the experimental data. 
While there remains a discrepancy between simulation and data of around 10-20\%, it might be explained by three observations. 
First, this analysis cuts requires only one jet in the final state which allows for contributions from the DIS region to leak into this measurement, for example if the scattered beam electron has not been correctly detected or identified. 
Secondly, the precise value of the strong coupling in the fit of the photon PDFs is not explicitly mentioned in the corresponding publications~\cite{Schuler1995,Schuler1996a}. An updated photon PDF fit would be performed with the current world average of $\alpha_S$ and might further reduce the discrepancy. 
Lastly, the modelling of multiple-parton interactions for photon--\-proton interactions needs a fitting to the data for both proton--\-proton and photon--\-proton data. 
As neither have been tuned to data yet it can be suspected that the MPI will receive larger contributions, thereby improving overall agreement of simulation and data. 

\begin{figure}[h!]
    \centering
    \begin{tabular}{ccc}
        \includegraphics[width=.3\linewidth]{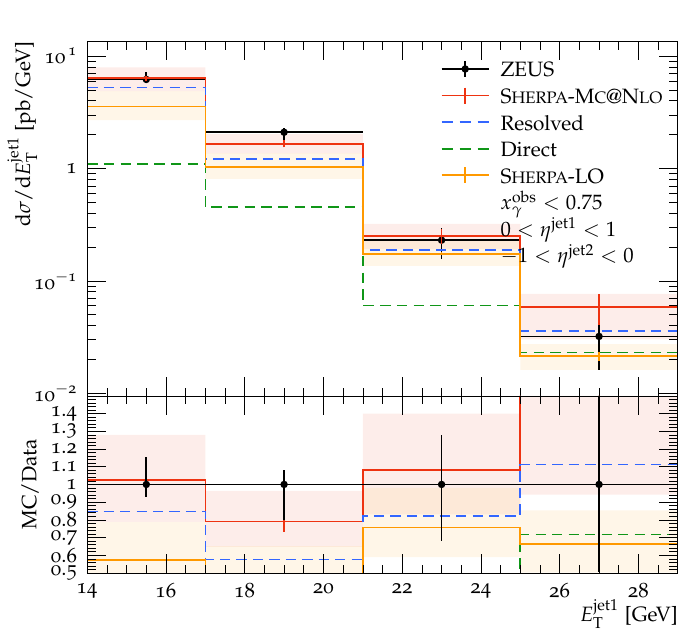} &
        \includegraphics[width=.3\linewidth]{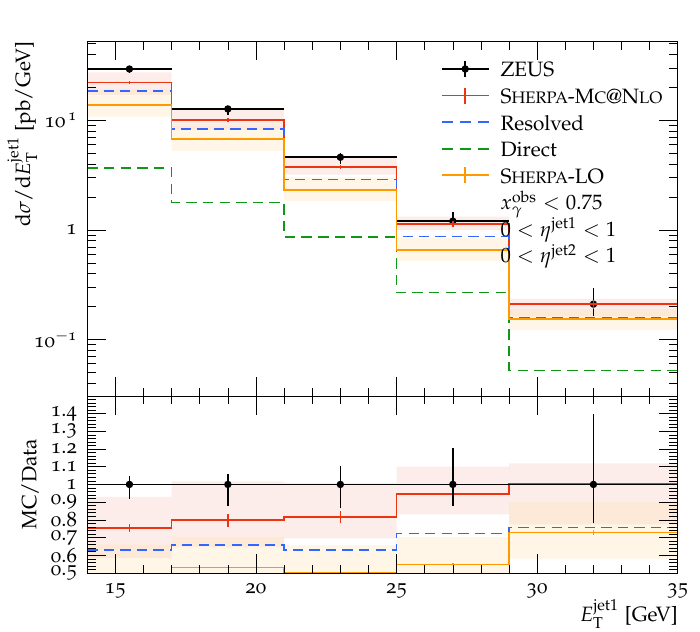} & \\ 
        \includegraphics[width=.3\linewidth]{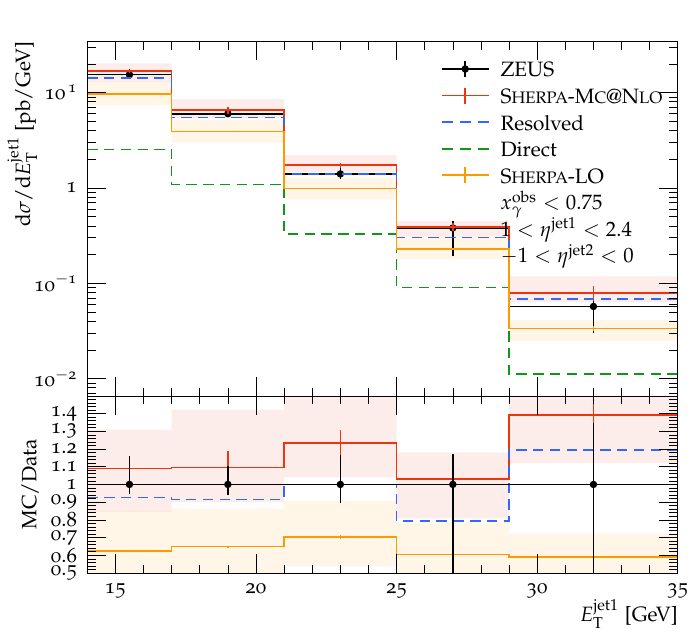} &
        \includegraphics[width=.3\linewidth]{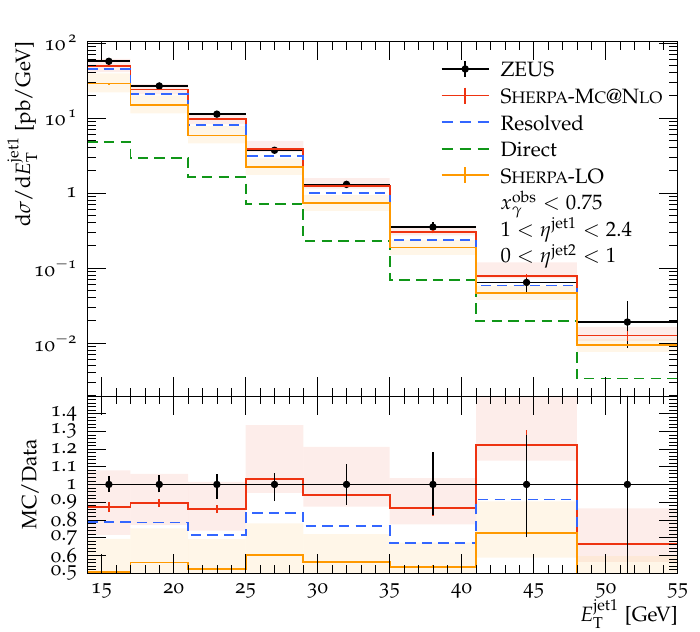} &
        \includegraphics[width=.3\linewidth]{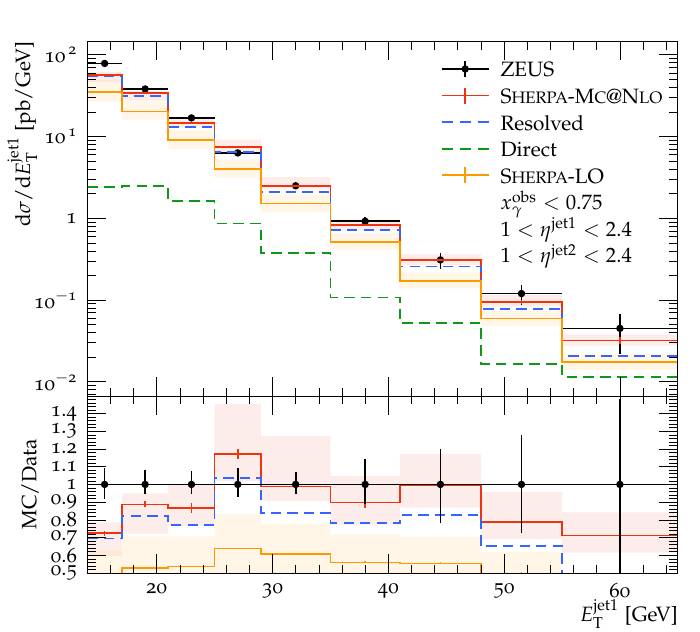}
    \end{tabular}
    \caption{Differential dijet inclusive cross section with respect to the transverse momentum of the leading jet, with \protect$0<\eta^{(1)}<1$ (upper panel) and \protect$1<\eta^{(1)}<2.4$ (lower panel) and in different bins for $\eta^{(2)}$, comparing results of our \protect\Sherpa \MCatNLO simulation at hadron-level incl. MPI effects with the LO simulation and with data~\protect\cite{ZEUS:2001zoq} taken by \protect\ZEUS at \protect\HERA Run 1.}
    \label{fig:nlo-zeus-2002b}
\end{figure}

The spectrum of the jet transverse energies $E_T$, displayed in Fig.\  ~\ref{fig:nlo-zeus-2012_2} exhibits a slight shape in the distribution for forward jets and small $E_T$. The same effect can be seen in Fig. \ref{fig:nlo-zeus-2002b}, where the simulation describes data well for central leading jets, but worsens as both go more forward. As this part of the phase receives contributions from the MPIs, we again suspect that the naive parameter choice underestimate the amount of additionally generated radiation. 

\subsection{Photon PDF for precision phenomenology}\label{Sec:photon-pdf-quality}

\begin{figure}[h!]
\centering
\begin{tabular}{cc}
    \includegraphics[width=.4\linewidth]{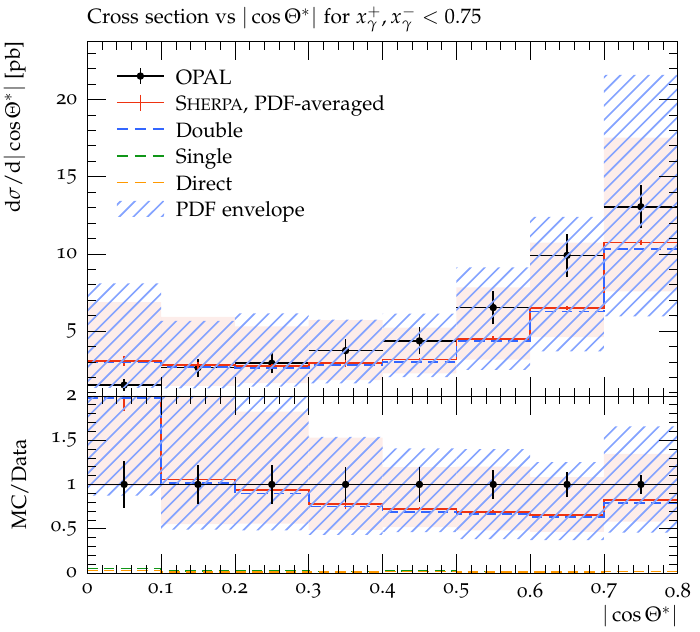} &
    \includegraphics[width=.4\linewidth]{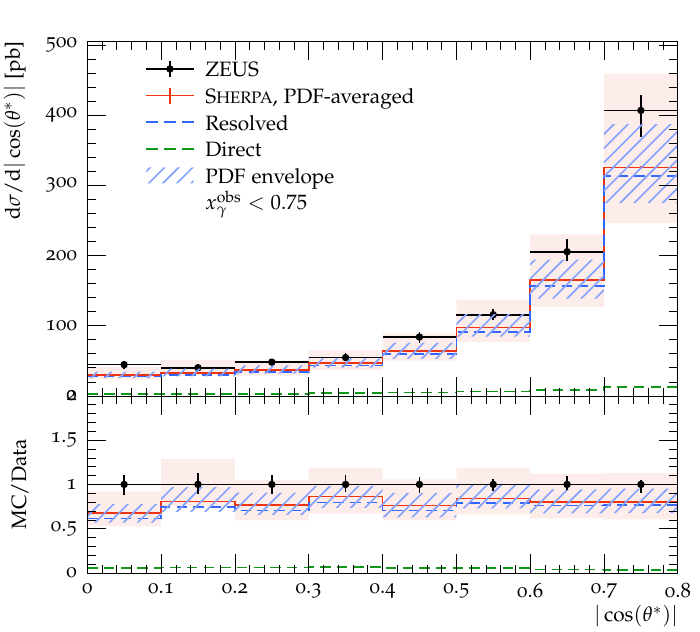} 
  \end{tabular}
  \caption{
  Distributions for $\cos\Theta^*$ at \LEP (left) and for $\cos\theta^*$ at \HERA (right), comparing \protect\Sherpa's LO simulations with data from \protect\OPAL~\protect\cite{OPAL:2003hoh} and \protect\ZEUS~\protect\cite{ZEUS:2001zoq}.  
  Scale uncertainties at LO are indicated by the pink band, while PDF uncertainties are shown with the blue hatched area.}
  \label{fig:pdf_uncerts}
\end{figure}
The predictions of photo-production cross sections and distributions in low-$x_\gamma$ space exhibit large variations depending on the used photon PDF. 
In fact, these deviations can be as large or even larger in value than the estimate of higher-order corrections through the scale variations, especially for the simulation of photo-production at lepton colliders, see Fig.\ \ref{fig:pdf_uncerts}. 
There we show the very inclusive $\cos\Theta^*$ and $\cos\theta^*$ distributions, obtained at leading order, at \LEP and \HERA, respectively., indicating LO scale and PDF uncertainties separately. 
For the latter we present the full range of results from all available leading order PDFs, and we observe that the PDF uncertainties alone are of equal size as the LO scale uncertainties.
This underlines the need for a comprehensive retuning of the photon PDFs with available data, and including higher-order calculations and simulations.

It is also worth noting that a consistent simulation of photo-production at hadron--\-lepton colliders necessitates a combined fit of the photon and proton PDFs. 
Depending on the proton PDF and its value for $\alpha_S$ the inclusive cross-section in a \HERA Run 2 simulation gives deviations of about 20\%, as can be seen in Tab.~\ref{tab:xs-proton-pdf-variation}. 
This underlines the necessity for a systematic refit of photon PDFs to use in conjunction with modern proton PDFs. 
While no new data has been taken since the retiring of the \HERA collider, a consistent fit to all the data with the updated values for $\alpha_S$ and including error estimates should increase the confidence in precision phenomenology for photo-production for the planned \EIC and other future colliders.
\begin{table}[ht!]
    \begin{center}
        \begin{tabular}{|l||c|c|}
        \hline
        $\vphantom{\frac{|^|}{|_|}}$ PDF & NNPDF23\_lo\_as\_0130\_qed & PDF4LHC21\_40\_pdfas \\
        $\vphantom{\frac{|^|}{|_|}}$ $\alpha_S$ & 0.13 & 0.118 \\
        $\vphantom{\frac{|^|}{|_|}}$ Order & 1 & 3 \\
        \hline
        $\vphantom{\frac{|^|}{|_|}}$ $\sigma (\gamma j \to j j)$ / nb & $2.85 \pm 0.02$ & $2.303 \pm 0.016$ \\
        $\vphantom{\frac{|^|}{|_|}}$ $\sigma (j j \to j j)$ / nb & $2.151 \pm 0.002$ & $1.997 \pm 0.002$ \\
        \hline
        \end{tabular}
        \parbox{0.8\textwidth}{\caption{
            Inclusive cross-sections for one million events in a \HERA Run 1 dijet photo-production setup with two different proton PDFs and the same PDF for the photon.}
            \protect\label{tab:xs-proton-pdf-variation}
        }
    \end{center}
\end{table}

\section{Summary}\label{Sec:summary}

In this paper we reported on developments of the \Sherpa event generator towards the first hadron-level simulation of photo-production at NLO accuracy, to be published in our upcoming \Sherpa 3.0 public release.
We described how the initial state phase space is treated to allow for a flexible yet efficient integration of different spectra and distributions, based on the equivalent photon approximation and the inclusion of a broad range of photon PDFs.
We validated our model by comparison to data from the \LEP and \HERA experiments and found satisfactory agreement between the data and the simulation, noting that the non-perturbative aspects of the simulation, in particular intrinsic $k_T$ and multiple-parton interactions, still require a comprehensive tuning to data.
Apart from possible improvement in this sector of the simulation, our analysis of uncertainties underlined the necessity for a refitting of the photon PDFs, especially in view of the anticipated precision of the planned electron-ion collider.
We believe that this step -- a first refitting of photon PDFs after 20 years -- based on higher-order calculations, modern simulation tools at least at NLO accuracy, and recent proton PDFs will siginifcantly improve the quality of our theoretical preprations for this new collider experiment.

\section*{Acknowledgements}
We are indebted to our colleagues in the \Sherpa collaboration, for numerous discussions and technical support.
F.K.\ gratefully acknowledges funding as Royal Society Wolfson Research fellow. 
F.K.\ and P.M.\ are supported by the STFC under grant agreement ST/P001246/1.
S.H.\ is supported by Fermi Research Alliance, LLC under Contract No. DE-AC02-07CH11359 with the U.S. Department of Energy, Office of Science, Office of High Energy Physics.
P.M.\ would like to thank the Fermilab Theory Division for hospitality while this work was finalized.

\bibliographystyle{unsrt}
\bibliography{journal,photon-pdfs,analyses}

\begin{thebibliography}{10}

\bibitem{BaBar:2018zpn}
J.~P. Lees et~al.
\newblock {Measurement of the $\gamma^{\star}\gamma^{\star} \to \eta'$
  transition form factor}.
\newblock {\em Phys. Rev. D}, 98(11):112002, 2018.

\bibitem{Ragoni:2021hmr}
Simone Ragoni.
\newblock {New measurements on diffractive vector mesons}.
\newblock {\em PoS}, LHCP2021:085, 2021.

\bibitem{Klein:2021syg}
Spencer~R. Klein.
\newblock {$\rho$ photoproduction in ALICE}.
\newblock In {\em {Low-x Workshop 2021}}, 12 2021.

\bibitem{Tu:2020mvm}
Zhoudunming Tu.
\newblock {Exclusive J/ photoproduction off deuteron in d+Au ultra-peripheral
  collisions at STAR}.
\newblock {\em PoS}, HardProbes2020:100, 2021.

\bibitem{OPAL:2003hoh}
G.~Abbiendi et~al.
\newblock {Dijet production in photon-photon collisions at s(ee)**(1/2) from
  189-GeV to 209-GeV}.
\newblock {\em Eur. Phys. J. C}, 31:307--325, 2003.

\bibitem{OPAL:2007jeb}
G.~Abbiendi et~al.
\newblock {Inclusive Jet Production in Photon-Photon Collisions at s(ee)**(1/2)
  from 189 to 209-GeV}.
\newblock {\em Phys. Lett. B}, 658:185--192, 2008.

\bibitem{L3:2004ehh}
P.~Achard et~al.
\newblock {Inclusive jet production in two-photon collisions at LEP}.
\newblock {\em Phys. Lett. B}, 602:157--166, 2004.

\bibitem{OPAL:1999pnw}
G.~Abbiendi et~al.
\newblock {Total hadronic cross-section of photon-photon interactions at LEP}.
\newblock {\em Eur. Phys. J. C}, 14:199--212, 2000.

\bibitem{OPAL:1998ggd}
K.~Ackerstaff et~al.
\newblock {Inclusive production of charged hadrons and K0(S) mesons in
  photon-photon collisions}.
\newblock {\em Eur. Phys. J. C}, 6:253--264, 1999.

\bibitem{OPAL:1999peo}
G.~Abbiendi et~al.
\newblock {Inclusive production of D*+- mesons in photon-photon collisions at
  S**(1/2)(ee) = 183-GeV and 189-GeV and a first measurement of F**gamma(2,c)}.
\newblock {\em Eur. Phys. J. C}, 16:579--596, 2000.

\bibitem{ZEUS:2001zoq}
S.~Chekanov et~al.
\newblock {Dijet photoproduction at HERA and the structure of the photon}.
\newblock {\em Eur. Phys. J. C}, 23:615--631, 2002.

\bibitem{ZEUS:2012pcn}
H.~Abramowicz et~al.
\newblock {Inclusive-jet photoproduction at HERA and determination of alphas}.
\newblock {\em Nucl. Phys. B}, 864:1--37, 2012.

\bibitem{ZEUS:1996uid}
M.~Derrick et~al.
\newblock {Measurement of the F2 structure function in deep inelastic e+ p
  scattering using 1994 data from the ZEUS detector at HERA}.
\newblock {\em Z. Phys. C}, 72:399--424, 1996.

\bibitem{ZEUS:1998agx}
J.~Breitweg et~al.
\newblock {ZEUS results on the measurement and phenomenology of F(2) at low x
  and low Q**2}.
\newblock {\em Eur. Phys. J. C}, 7:609--630, 1999.

\bibitem{ZEUS:2002wfj}
S.~Chekanov et~al.
\newblock {Exclusive photoproduction of J / psi mesons at HERA}.
\newblock {\em Eur. Phys. J. C}, 24:345--360, 2002.

\bibitem{H1:2003xoe}
C.~Adloff et~al.
\newblock {Measurement and QCD analysis of neutral and charged current
  cross-sections at HERA}.
\newblock {\em Eur. Phys. J. C}, 30:1--32, 2003.

\bibitem{H1:2000kis}
C.~Adloff et~al.
\newblock {Elastic photoproduction of J / psi and Upsilon mesons at HERA}.
\newblock {\em Phys. Lett. B}, 483:23--35, 2000.

\bibitem{Glueck1992}
M.~Glück, E.~Reya, and A.~Vogt.
\newblock Photonic parton distributions.
\newblock {\em Physical Review D}, 46(5):1973--1979, September 1992.

\bibitem{Cornet:2002iy}
F.~Cornet, P.~Jankowski, M.~Krawczyk, and A.~Lorca.
\newblock {A New five flavor LO analysis and parametrization of parton
  distributions in the real photon}.
\newblock {\em Phys. Rev. D}, 68:014010, 2003.

\bibitem{Cornet:2004nb}
F.~Cornet, P.~Jankowski, and M.~Krawczyk.
\newblock {A New 5 flavor NLO analysis and parametrizations of parton
  distributions of the real photon}.
\newblock {\em Phys. Rev. D}, 70:093004, 2004.

\bibitem{Slominski2005}
W.~Slominski, H.~Abramowicz, and A.~Levy.
\newblock {NLO photon parton parametrization using ee and ep data}.
\newblock {\em Eur.Phys.J.C45:633-641,2006}, 45, April 2005.

\bibitem{Schuler1995}
Gerhard~A. Schuler and Torbjörn Sjöstrand.
\newblock Low- and high-mass components of the photon distribution functions.
\newblock {\em Zeitschrift für Physik C Particles and Fields}, 68(4):607--623,
  December 1995.

\bibitem{Schuler1996a}
Gerhard~A. Schuler and Torbjörn Sjöstrand.
\newblock {Parton Distributions of the Virtual Photon}.
\newblock {\em Phys.Lett.B376:193-200,1996}, January 1996.

\bibitem{Schuler:1996en}
Gerhard~A. Schuler and Torbj{\"o}rn Sj{\"o}strand.
\newblock {A scenario for high-energy $\gamma \gamma$ interactions}.
\newblock {\em Z. Phys.}, C73:677--688, 1997.

\bibitem{Sjostrand:2006za}
Torbj{\"o}rn Sj{\"o}strand, Stephen Mrenna, and Peter Skands.
\newblock {PYTHIA 6.4 physics and manual}.
\newblock {\em JHEP}, 05:026, 2006.

\bibitem{Bierlich:2022pfr}
Christian Bierlich et~al.
\newblock {A comprehensive guide to the physics and usage of PYTHIA 8.3}.
\newblock {\em SciPost Phys. Codebases}, 3 2022.

\bibitem{Klasen:1996it}
M.~Klasen and G.~Kramer.
\newblock {Inclusive two jet production at HERA: Direct and resolved
  cross-sections in next-to-leading order QCD}.
\newblock {\em Z. Phys. C}, 76:67--74, 1997.

\bibitem{Frixione:1997ks}
Stefano Frixione and Giovanni Ridolfi.
\newblock {Jet photoproduction at HERA}.
\newblock {\em Nucl. Phys.}, B507:315--333, 1997.

\bibitem{Kramer:1998bc}
G.~Kramer and B.~P{\"o}tter.
\newblock {Low $Q^2$ jet production at HERA in next-to-leading order QCD}.
\newblock {\em Eur. Phys. J.}, C5:665--679, 1998.

\bibitem{Klasen:1997br}
M.~Klasen, T.~Kleinwort, and G.~Kramer.
\newblock {Inclusive Jet Production in $\gamma p$ and $\gamma\gamma$ Processes:
  Direct and Resolved Photon Cross Sections in Next-To-Leading Order QCD}.
\newblock {\em Eur. Phys. J. direct}, 1(1):1, 1998.

\bibitem{Harris:1997hz}
B.~W. Harris and J.~F. Owens.
\newblock {Photoproduction of jets at HERA in next-to-leading order QCD}.
\newblock {\em Phys. Rev. D}, 56:4007--4016, 1997.

\bibitem{Guzey:2023syh}
V.~Guzey and M.~Klasen.
\newblock {Inclusive and Diffractive Dijet Photoproduction at the
  Electron\textendash{}Ion Collider in NLO QCD}.
\newblock {\em Acta Phys. Polon. Supp.}, 16(7):7--A11, 2023.

\bibitem{Chang:2009uj}
Chao-Hsi Chang, Rong Li, and Jian-Xiong Wang.
\newblock {J/psi polarization in photo-production up-to the next-to-leading
  order of QCD}.
\newblock {\em Phys. Rev. D}, 80:034020, 2009.

\bibitem{Eskola:2022vpi}
Kari~J. Eskola, Christopher~A. Flett, Vadim Guzey, Topi L\"oyt\"ainen, and
  Hannu Paukkunen.
\newblock {Exclusive J/\ensuremath{\psi} photoproduction in ultraperipheral
  Pb+Pb collisions at the CERN Large Hadron Collider calculated at
  next-to-leading order perturbative QCD}.
\newblock {\em Phys. Rev. C}, 106(3):035202, 2022.

\bibitem{Eskola:2022vaf}
Kari~J. Eskola, Christopher~A. Flett, Vadim Guzey, Topi L\"oyt\"ainen, and
  Hannu Paukkunen.
\newblock {Next-to-leading order perturbative QCD predictions for exclusive
  J/\ensuremath{\psi} photoproduction in oxygen-oxygen and lead-lead collisions
  at energies available at the CERN Large Hadron Collider}.
\newblock {\em Phys. Rev. C}, 107(4):044912, 2023.

\bibitem{Eskola:2023oos}
Kari~J. Eskola, Christopher~A. Flett, Vadim Guzey, Topi L\"oyt\"ainen, and
  Hannu Paukkunen.
\newblock {Predictions for exclusive $\varUpsilon $ photoproduction in
  ultraperipheral ${\textrm{Pb}}+{\textrm{Pb}}$ collisions at the LHC at
  next-to-leading order in perturbative QCD}.
\newblock {\em Eur. Phys. J. C}, 83(8):758, 2023.

\bibitem{Jones:2015nna}
S.~P. Jones, A.~D. Martin, M.~G. Ryskin, and T.~Teubner.
\newblock {Exclusive $J/\psi$ and $\Upsilon$ photoproduction and the low $x$
  gluon}.
\newblock {\em J. Phys. G}, 43(3):035002, 2016.

\bibitem{Jones:2013pga}
S.~P. Jones, A.~D. Martin, M.~G. Ryskin, and T.~Teubner.
\newblock {Probes of the small $x$ gluon via exclusive $J/\psi$ and $\Upsilon$
  production at HERA and the LHC}.
\newblock {\em JHEP}, 11:085, 2013.

\bibitem{Gleisberg:2008ta}
T.~Gleisberg, S.~H{\"o}che, F.~Krauss, M.~Sch\"{o}nherr, S.~Schumann,
  F~Siegert, and J.~Winter.
\newblock {Event generation with \Sherpa 1.1}.
\newblock {\em JHEP}, 02:007, 2009.

\bibitem{Bothmann:2019yzt}
Enrico Bothmann et~al.
\newblock {Event Generation with Sherpa 2.2}.
\newblock {\em SciPost Phys.}, 7(3):034, 2019.

\bibitem{Kleiss:1994qy}
Ronald Kleiss and Roberto Pittau.
\newblock {Weight optimization in multichannel Monte Carlo}.
\newblock {\em Comput. Phys. Commun.}, 83:141--146, 1994.

\bibitem{vonWeizsacker:1934sx}
C.F. von Weizs{\"a}cker.
\newblock {Radiation emitted in collisions of very fast electrons}.
\newblock {\em Z.Phys.}, 88:612--625, 1934.

\bibitem{Williams:1934ad}
E.J. Williams.
\newblock {Nature of the high-energy particles of penetrating radiation and
  status of ionization and radiation formulae}.
\newblock {\em Phys.Rev.}, 45:729--730, 1934.

\bibitem{Budnev:1974de}
V.~M. Budnev, I.~F. Ginzburg, G.~V. Meledin, and V.~G. Serbo.
\newblock {The two photon particle production mechanism. Physical problems.
  Applications. Equivalent photon approximation}.
\newblock {\em Phys. Rept.}, 15:181--281, 1974.

\bibitem{Frixione:1993yw}
Stefano Frixione, Michelangelo~L. Mangano, Paolo Nason, and Giovanni Ridolfi.
\newblock {Improving the Weizsacker-Williams approximation in electron - proton
  collisions}.
\newblock {\em Phys. Lett. B}, 319:339--345, 1993.

\bibitem{sherpa-manual}
{Sherpa Project webpage}.
\newblock \url{https://sherpa-team.gitlab.io/}.
\newblock Accessed: 13 Oct 2023.

\bibitem{Frixione:1997ma}
Stefano Frixione, Michelangelo~L. Mangano, Paolo Nason, and Giovanni Ridolfi.
\newblock {Heavy quark production}.
\newblock {\em Adv. Ser. Direct. High Energy Phys.}, 15:609--706, 1998.

\bibitem{Catani:1996vz}
S.~Catani and M.~H. Seymour.
\newblock {A general algorithm for calculating jet cross sections in NLO QCD}.
\newblock {\em Nucl. Phys.}, B485:291--419, 1997.

\bibitem{Krauss:2001iv}
Frank Krauss, Ralf Kuhn, and Gerhard Soff.
\newblock {AMEGIC++ 1.0: A Matrix Element Generator In C++}.
\newblock {\em JHEP}, 02:044, 2002.

\bibitem{Gleisberg:2008fv}
Tanju Gleisberg and Stefan H{\"o}che.
\newblock {Comix, a new matrix element generator}.
\newblock {\em JHEP}, 12:039, 2008.

\bibitem{Gleisberg:2007md}
Tanju Gleisberg and Frank Krauss.
\newblock {Automating dipole subtraction for QCD NLO calculations}.
\newblock {\em Eur. Phys. J.}, C53:501--523, 2008.

\bibitem{Buccioni:2019sur}
Federico Buccioni, Jean-Nicolas Lang, Jonas~M. Lindert, Philipp Maierh{\"o}fer,
  Stefano Pozzorini, Hantian Zhang, and Max~F. Zoller.
\newblock {OpenLoops 2}.
\newblock {\em Eur. Phys. J. C}, 79(10):866, 2019.

\bibitem{Schumann:2007mg}
Steffen Schumann and Frank Krauss.
\newblock {A parton shower algorithm based on Catani-Seymour dipole
  factorisation}.
\newblock {\em JHEP}, 03:038, 2008.

\bibitem{Frixione:2002ik}
Stefano Frixione and Bryan~R. Webber.
\newblock {Matching NLO QCD computations and parton shower simulations}.
\newblock {\em JHEP}, 06:029, 2002.

\bibitem{Frixione:2003ei}
S.~Frixione, P.~Nason, and B.~R. Webber.
\newblock {Matching NLO QCD and parton showers in heavy flavour production}.
\newblock {\em JHEP}, 08:007, 2003.

\bibitem{Hoeche:2011fd}
Stefan H{\"o}che, Frank Krauss, Marek Sch{\"o}nherr, and Frank Siegert.
\newblock {A critical appraisal of NLO+PS matching methods}.
\newblock {\em JHEP}, 09:049, 2012.

\bibitem{Sjostrand:1987su}
Torbj{\"o}rn Sj{\"o}strand and Maria van Zijl.
\newblock {A multiple-interaction model for the event structure in hadron
  collisions}.
\newblock {\em Phys. Rev.}, D36:2019, 1987.

\bibitem{Schuler:1993wr}
Gerhard~A. Schuler and Torbjorn Sjostrand.
\newblock {Hadronic diffractive cross-sections and the rise of the total
  cross-section}.
\newblock {\em Phys. Rev. D}, 49:2257--2267, 1994.

\bibitem{Chahal:2022rid}
Gurpreet~Singh Chahal and Frank Krauss.
\newblock {Cluster Hadronisation in Sherpa}.
\newblock {\em SciPost Phys.}, 13(2):019, 2022.

\bibitem{Bierlich:2019rhm}
Christian Bierlich et~al.
\newblock {Robust Independent Validation of Experiment and Theory: Rivet
  version 3}.
\newblock {\em SciPost Phys.}, 8:026, 2020.

\end{thebibliography}
\end{document}